\documentclass[apj,iop]{emulateapj}
\usepackage{natbib}
\usepackage{graphicx}

\slugcomment{Submitted to the Astrophysical Journal on August 11, 2011}

\shorttitle{The Period-Mass-Radius Distribution of Super-Earths/Sub-Neptunes}
\shortauthors{Wolfgang \& Laughlin}

\begin{document}


\title{Combining \emph{Kepler} and HARPS Occurrence Rates to Infer the Period-Mass-Radius Distribution of Super-Earths/Sub-Neptunes}


\author{Angie Wolfgang\altaffilmark{1} and Gregory Laughlin}
\affil{Department of Astronomy and Astrophysics, \\University of California,
    Santa Cruz, CA 95064}
\altaffiltext{1}{Eugene Cota-Robles Fellow, UCSC; NSF Graduate Research Fellow}
\email{wolfgang@ucolick.org}



\begin{abstract}

The ongoing High Accuracy Radial velocity Planet Search (HARPS) has found that 30-50\% of GK dwarfs in the solar neighborhood host planets with $M_{pl}\lesssim$ M$_{Nep}$ in orbits of $P\leq50$ days.  At first glance, this overall occurrence rate seems inconsistent with the planet frequency measured during Q0-Q2 of the \emph{Kepler Mission}, whose 1,235 detected planetary candidates imply that $\sim$15\% of main sequence dwarfs harbor short-period planets with $R_{pl}<4$ R$_\oplus$.  A rigorous comparison between the two surveys is difficult, however, as they observe different stellar populations and measure different planetary properties.  Here we report the results of a Monte Carlo study that can account for this discrepancy via plausible distributions of planetary compositions.  We find that a population concurrently consisting of (1) dense silicate-iron planets and (2) low-density gas-dominated worlds provides a natural fit to the current data.  In this scenario, the fraction of dense planets decreases with increasing mass, from $f_{rocky}=90\%$ at $M=1$ M$_\oplus$ to $f_{rocky}=10\%$ at $M=$ M$_{Nep}$.  Our best fit population has a total occurrence rate of 40\% for $2\leq P\leq50$ days and $1\leq M\leq17$ M$_\oplus$, and is characterized by simple power-law indices of the form $N(M)dM\propto M^\alpha dM$ and $N(P)dP\propto P^\beta dP$ with $\alpha=-1.0$ and $\beta=0.0$.  Our model population therefore contains four free parameters and is readily testable with future observations.  Furthermore, our model's insistence that at least two distinct types of planets must exist in the survey data indicates that multiple formation mechanisms are at work to produce the population of planets commonly referred to as ``super-Earths".

\end{abstract}


\section{Introduction}

The past decade has seen an extraordinary increase in our understanding of short-period extrasolar planets, due in large part to the plethora of transit and radial velocity (RV) detections made by a host of planet search surveys.  Now with over 500 planets known \citep{Wri10, Schn11}, the state of the art has progressed to the point where statistical studies of entire planet populations are realistically feasible \citep{Cum08, How10, For11, Lat11, Moo11, How11,Wit11,You11,Tre11}.  However, a substantial challenge still lies in synthesizing the results from different surveys into a cohesive picture of the Galactic planetary population, as each technique provides different information about the planets' physical characteristics and is subject to different selection biases.

These cross-survey considerations are especially important when one tries to compare the results of Doppler velocity surveys with the results of photometric transit surveys.  Not only do these two detection methods generally sample different regions of the Galaxy, but they also implement different observing strategies due to the intrinsically low geometric probability of a planetary transit and to the strict spectroscopic requirements needed to achieve 1 m/s precision in RV (see for example \citealp{Bor10, Koc10, Bat10,Rup04}).  The result of these fundamental differences is that most RV-detected planets don't transit, and that most transiting planets suffer from a dearth of high-precision Doppler follow-up measurements.  All is not lost, however: if these biases are properly accounted for, then one can utilize the statistical properties of the two samples to draw conclusions about the Galactic distribution of planetary properties.

A particularly valuable outcome of the transit-RV comparison originates from the distinction between measuring a planet's radius via a transit and measuring its mass via RV observations.  Because of this difference, when a Doppler-characterized planet is observed to transit, its range of possible compositions can be modeled even in the absence of any other observational constraints, through individual mass-to-radius (M-R) relationships calculated for a variety of interior planetary structures (e.g. \citealp{For07, Sea07, Rog11}).  Unfortunately, however, planets that are well-characterized by both methods are rare, necessitating the use of statistical techniques to make some headway in understanding the compositional distribution of planet populations if we assume that transit and RV surveys adequately sample the full range and frequency of planetary compositions once the populations are corrected for selection bias.

Before these population-wide M-R relationships can be properly interpreted, it is essential to understand how they are fundamentally different from the M-Rs that are calculated using structural models of individual planets.  On the most basic level, the transformation of a \emph{population} of planetary masses to radii requires that a range of compositions be included \emph{a priori}.  While it is certainly true that inferring an individual planet's composition from its mass and radius is a degenerate problem and \emph{results} in a range of possible part-iron, part-silicate, part-gas compositions, the bulk density of a planet, $\rho(M,R)$, is nonetheless a deterministic quantity.  This connection is absent, however, when one compares a transiting planet population to a Doppler-detected population and the two samples have very few planets in common.  As a result, bulk density is essentially a free parameter in transit-RV comparisons, and some assumptions about it, or about the compositions which correspond to it, must be made.  

A key issue for the transit-RV comparison is how one chooses to parameterize planetary composition over the entire population.  The simplest case would be if all planets had the same composition, as this enables the planets' masses to be straightforwardly converted to radii.  However, \citet{How11} have already shown that this very simple M-R fails to match the \emph{Kepler} planet candidates when a power law is used for the planetary mass distribution, and so we consider more flexible and more physically motivated M-Rs in this paper.  In particular, we find that the choice of these compositional parameters is crucial for reconciling any apparent statistical disparities between RV and transiting planet populations.

The plethora of ongoing planet searches enables the Galactic planetary census to be illuminated in a number of different ways.  Two of the most influential surveys to date are the \emph{Kepler Mission}, which found 1,235 transiting planet candidates in its first four months of data \citep{Bor11}, and the Geneva High Accuracy Radial velocity Planet Search (HARPS), which has discovered over 85 planets around hundreds of the brightest stars in the solar neighborhood \citep{Seg11}.  Both of these surveys are in a position to unearth the population of low-mass short-period planets and to provide statistics about their relative frequency, which hints suggestively at the prevalence of truly Earth-like planets and which is of particular interest for planet formation theories that strive to explain or predict the mass-distance distribution of planet populations \citep{Ida04, Kor06, Sch09, Mor09, Ida10, Ali11}.  

Alarmingly, the low-mass planet occurrence rates measured by the two surveys appear to conflict with one another.  Systematic statistical analyses of the short-period \emph{Kepler} planet candidates have yielded $0.130 \pm 0.008$ planets per solar-type star \citep{How11} or 0.19 planets per solar-type star \citep{You11}, with the planets having $2 \leq R_{pl} \leq 4$ R$_\oplus$ and $P \leq 50$ days.  On the other hand, preliminary results from the HARPS planet search \citep{Lov09,May09,Udr10} indicate that 30 - 50\% of Sun-like stars host sub-Neptune mass planets within 50-day orbits --- a planet frequency that is substantially higher than the \emph{Kepler} occurrence rate.  Although these two occurrence rates do provide somewhat different information as discussed in \S \ref{Discuss}, the following order-of-magnitude argument readily gives a sense for the apparent discrepancy in terms of the total number of planets that \emph{Kepler} would have detected in its first four months of data.  

Given a 40\% occurrence rate and $\sim$ 150,000 \emph{Kepler} target stars, there are 60,000 potentially detectable planets in \emph{Kepler}'s field of view, assuming that each host star harbors only one planet.  Not all of these planets will transit, however, as the required star-planet-observer alignment is fairly improbable given random inclinations along the line of sight.  For planets in orbits of 50 days or less, this geometrical transit probability works out to be $1-15\%$; taking a 5\% transit probability (10-day orbit) as a benchmark, the number of sub-Neptune-mass planets that \emph{Kepler} would have been able to detect is thus approximately 3,000.  If we map the \emph{Kepler} planet candidate radii to mass via the simple relation $M/M_\oplus = (R/R_\oplus)^{2.06}$ \citep{Lis11b}, then we see that about 900 of \emph{Kepler}'s planet candidates fall in the $M <$ M$_{Nep}$ range.  Thus, the HARPS occurrence rate appears to overestimate the number of planets that \emph{Kepler} would have detected by a factor of 3.

Order-of-magnitude arguments can be misleading, however, so in this paper we take care to fully account for details of the RV-transit comparison that may affect this result, including factors such as the enhanced geometrical transit probability of elliptical orbits, the shallower transits of more inclined orbits due to stellar limb darkening, target star selection biases, and \emph{Kepler}'s detection incompleteness.   By conducting a detailed comparison between the second quarter \emph{Kepler} planet candidates and the HARPS's generalized statistic about the occurrence rate of low-mass planets around solar-type stars, we provide a systematic statistical analysis of the compositions in a truly sub-Neptune-mass exoplanet population.  In the process, we identify the first physically motivated mass-to-radius relationship for a population of low-mass, short-period planets that can reproduce occurrence rates observed by both RV and transit planet searches.  

The layout of this paper is as follows.  In \S \ref{Kepdata} we briefly summarize the \emph{Kepler} data set and discuss the use of planet candidates instead of confirmed planets for our analysis.  In \S \ref{Sims} we describe our simulations and statistical calculations.  In \S \ref{Results} we present our results on the total number of planets that \emph{Kepler} would have been able to detect given the HARPS occurrence rate, and in \S \ref{Discuss} we discuss the implications of these results for our current understanding of exoplanet populations.

\section{The \emph{Kepler} Data Set} \label{Kepdata}

The \emph{Kepler Mission} is a 3.5-year search for potentially habitable Earth-sized planets around solar-type stars \citep{Bor10}.  To detect these planets, \emph{Kepler} monitors $\sim$ 150,000 stars in its $> 100$ deg$^2$ field of view \citep{Koc10} for periodic photometric dips that fit the shape and duration of a planetary transit.  The telescope's $\sim 0.01^{\prime\prime}$ per hour pointing stability \citep{Koc10} and 10-100 ppm photometric precision \citep{Jen10} enables the detection of planets that are Earth-sized or smaller.  Furthermore, its Earth-trailing heliocentric orbit facilitates continuous data acquisition without the diurnal or annual cycles that generate aliases \citep{Koc10}.  These characteristics, along with the low false positive detection rate discussed below, enable robust statistical analysis of the detected planets' properties, including the small planetary radii and the short periods that are the focus of this study.  

On February 1, 2011 \emph{Kepler} released its second quarter (Q2) data, which was soon followed by the announcement of 1,235 transiting planet candidates \citep{Bor11}.  It is necessary to note, however, that the vast majority of these planets are unconfirmed and thus maintain ``planet candidate" status.  The current consensus is that these candidates can be catalogued as true planets only if they exhibit transit timing variations or are detected through the radial velocity method, as other astrophysical events such as binary blends with background stars, eclipsing hierarchical triples with small separations, and certain types of stellar variability can mimic planetary transits \citep{Gau10, Mor11}.  Unfortunately, however, the majority of \emph{Kepler}'s target stars have $V > 11$ and thus are faint for the purpose of Doppler follow-up, making these additional RV measurements expensive and leaving the vast majority of the \emph{Kepler} candidates unconfirmed.  

To compensate for these observational limitations, the \emph{Kepler} team has developed an extensive vetting process to eliminate as many of these false planetary transit signatures as possible \citep{Gau10, Bor11}.  Inevitably, however, a small but non-negligible fraction of false positives are expected to persist in the list of planet candidates.  \citet{Bor11} estimates that this false positive fraction is as high as 20\%, while a detailed Bayesian analysis conducted by \citet{Mor11} finds that the transit depth-independent false alarm probability is $< 5 \%$ over the entire field of view, given stars with \emph{Kepler} magnitude Kp $\leq 16$, a 30-50\% planet occurrence prior, and the assumption that follow-up astrometry can identify binaries at any Kp with separations $> 2^{\prime\prime}$.  When this last assumption is relaxed, the false alarm probability increases with decreasing transit depth for the fainter target stars and can exceed 15\%, as illustrated by the false positive probabilities that \citet{Mor11} individually compute for each \emph{Kepler} planet candidate.  We thus keep in mind that the total number of confirmed \emph{Kepler} planets is likely $ \sim 5 - 15\%$ lower than that reported by \citet{Bor11} as we proceed with our statistical analysis of the \emph{Kepler} planet candidate population.

\section{The Transit-RV Comparison: Methods} \label{Sims}

Comparing the HARPS occurrence rate with \emph{Kepler}'s planet candidates involves several steps.  First, we require that the aggregate properties of our initial planet population are consistent with the cumulative characteristics of the low-mass population detected by the HARPS survey (\S \ref{Define}).  To compare these planets to \emph{Kepler}'s public data set, we map our initial distribution of planet masses to radii via a population-wide mass-to-radius (M-R) relationship (\S \ref{MtoR}).  Each simulated planet is subsequently matched to a \emph{Kepler} target star (\S \ref{Selstar}) and its light curve is computed based on analytic transit formulae \citep{Man02}.  We then apply \emph{Kepler}'s detection criteria \citep{Jen10} to assess whether or not that planet would have been detected by the end of the second quarter (\S \ref{Detect}).  Finally, we use both the two-dimensional Kolmogorov-Smirnov test over period and radius and the one-dimensional $\chi^2$ test over radius to assess the quality of the fit between our simulated planet population and \emph{Kepler}'s planet candidates, and to identify the values of the free parameters which best fit \emph{Kepler}'s Q2 data (\S \ref{Stats}, \S \ref{Results}).

\subsection{Simulations of the Radial Velocity Population} \label{Define}


Other than stating that 30 - 50\% of Sun-like stars host sub-Neptune-mass planets with $P \le 50$ days, the HARPS overall occurrence rate does not address specific details of the planetary mass-period distribution.  Accordingly, we must select a general, easily parameterized distribution that is able to recover the HARPS overall occurrence rate.  Power laws meet these criteria, so we follow common practice \citep{Cum08, How10, How11, You11} and adopt:
\begin{equation}
N(M)dM = N_{tot} C_M M^\alpha dM,
\end{equation}
where $N(M)dM$ is the number of planets that have a mass between $M$ and $M + dM$, $N_{tot}$ is the total number of planets in the sample, $C_M$ is a normalization constant, and $\alpha$ is the mass power law index.  Similarly, we take for the period distribution:
\begin{equation}
N(P)dP = N_{tot} C_P P^\beta dP.
\end{equation}

We use the HARPS overall occurrence rate to determine $N_{tot}$, $C_M$, and $C_P$ for our simulated populations.  $N_{tot}$ is simply the planet occurrence rate times the total number of stars that \emph{Kepler} is observing, assuming that each star which harbors a planet harbors no more than that one planet---the bare minimum suggested by the prediction.  Given that \emph{Kepler} observed over 110,000 G and K dwarfs during its second quarter (Q2) of data \citep{KMT11}, this leads to $N_{tot} \sim 55000$ for a 50\% occurrence rate.  $C_M$ and $C_P$ are determined by setting minimum and maximum values for mass and period in our simulations.  The maximum period of 50 days is explicitly given by the stated HARPS occurrence rate, as are the limits on planet mass if we define a sub-Neptune planet to have $1 \le M \le 17$ M$_\oplus \sim$ M$_{Nep}$.  It is important to emphasize that we are only considering low planetary masses here; Jupiter-mass planets are not considered in our simulations.

The minimum value on period, while not expressly indicated in the HARPS low-mass occurrence rate, can be reasonably chosen from existing trends.  Both the census of \emph{Kepler} planet candidates \citep{Bor11} and the population of planets discovered through the radial velocity method \citep{Wri10} suggest that there is a dearth of planets with $P < 2$ days that is not due to the selection biases of the different detection methods.  \citet{How11} fit a power-law distribution with an exponential cutoff at short periods to the \emph{Kepler} planet candidates and found that the transitional period varies from 2 to 7 days for planets with $2 \leq R \leq 6$ R$_\oplus$.  To simplify our input distributions, we ignore the exponential cutoff and set $P_{min} = 2$ days, keeping in mind that any deviation from a power law for $2 \leq P \leq 7$ days may impact our ability to fit \emph{Kepler}'s observed distribution.

A rigorous interpretation of the HARPS statistic would include the unknown sin($i$) factor on the observed masses.  We note, however, that the distribution of inclinations for the observed radial velocity planets is poorly understood and that spherical isotropy cannot be assumed due to the detection biases inherent in the radial velocity technique.  Although some insight may be gleaned from statistical analysis such as that in \citet{Ho10} or from the few planets which exhibit the Rossiter-McLaughlin Effect \citep{Sch10}, in this analysis we take our mass limits as the bounds on the true mass of our simulated planets, effectively ignoring any refinements stemming from the sin($i$) effect.

\subsubsection{Simulation Parameters} \label{Simpar}

To account for the ambiguity in the RV mass and period distributions, we require that the power-law indices $\alpha$ and $\beta$ serve as free parameters in our simulations: we allow $\alpha$ to vary from -2.5 to 0 and $\beta$ to vary from -0.5 to 0.5, both in increments of 0.1.  We model eccentricity, $e$, longitude of periastron, $\omega$, and inclination, $i$, as uniform distributions, randomly drawing $e$ from $0 \leq e \leq 0.2$, $\omega$ from $0 \leq \omega < 2\pi$, and $i$ from an isotropic sphere.  Taken with $P$, these orbital elements serve to determine which planets transit, given their geometrical transit probability.  We choose to include non-zero eccentricities because elliptical orbits can enhance the probability of a transit, but we set the upper bound at $e = 0.2$ with the expectation that many short-period planets will have experienced a significant degree of tidal circularization.  This bound is broadly consistent with the observed eccentricity distribution of confirmed planets in our mass and period range, which shows that a vast majority ($\sim 80\%$) of low-mass planets with $P< 50$ days have $e \lesssim 0.2$ \citep{Wri10}.  

Two more free parameters are introduced for the second M-R we consider in this paper (\S \ref{multiMtoR}), as we allow the fraction of rocky planets in the population to vary as a linear function of mass.  These fractions are then used to randomly assign each planet either a gaseous or a rocky composition.  In addition, we randomly allocate each planet to a \emph{Kepler} target star, as discussed in \S \ref{Selstar}.

\subsection{Population-wide Mass-to-Radius Relationships} \label{MtoR}

A crucial consideration for the transit-RV comparison is the population-wide M-R used to map an RV planet's mass to a transiting planet's radius.  \citet{How11} have shown that applying one bulk density to an entire planet population fails to match the \emph{Kepler} candidates, so we begin our investigation with more flexible and physically motivated M-Rs, while taking care to minimize the number of degrees of freedom.  In particular, we consider two population-wide M-Rs in this paper: a power-law fit to measured planetary masses and radii, and a multi-valued parameterization that relaxes the single-valued assumption involved in fitting a power law to data.

\subsubsection{Single-Valued M-R} \label{singleMtoR}

\citet{Lis11b} use the following power-law fit to Earth and Saturn as the mass-radius relation for \emph{Kepler}'s planet candidates: 
\begin{equation} \label{LissMR}
\frac{M}{M_\oplus} = \left( \frac{R}{R_\oplus} \right) ^{2.06},
\end{equation}
which tacitly assumes that extrasolar planets resemble those in our Solar System.  Experience has shown that such an approach requires caution, so as a check we derive a comparable M-R for the nine transiting extrasolar planets with $1 \leq M \leq 17$ M$_\oplus$ \citep{Wri10}: CoRoT-7 b \citep{Que09, Leg09}, GJ 1214 b \citep{Char09}, Kepler-10 b \citep{Bat11}, Kepler-11 b - f \citep{Lis11a}, and 55 Cnc e \citep{Win11,Dem11}.  Including the error on radius, we find the following best fit for the radii of these planets given their masses:
\begin{equation} \label{exoPL1}
\frac{R}{R_\oplus} = 0.87^{+ 0.09}_{- 0.08} \left( \frac{M}{M_\oplus} \right) ^{0.45 \pm 0.06},
\end{equation}
which is encouragingly close to the inverse of the M-R used by \citet{Lis11b}.  However, the mass measurement errors are much larger than those on radius, so when we compute the fit in the other direction we find
\begin{equation}
\frac{M}{M_\oplus} = 5.8^{+ 1.2}_{- 1.0} \left( \frac{R}{R_\oplus} \right) ^{0.30 \pm 0.22},
\end{equation}
which has substantial error bars and does not invert to give Equation \ref{exoPL1}.  Thus, the M-R computed directly from the dually-detected, low-mass extrasolar planets is poorly constrained, and we proceed cautiously with $R/R_\oplus = (M/M_\oplus)^{0.48}$ from Equation \ref{LissMR}.

\subsubsection{Multi-Valued M-R} \label{multiMtoR}

While the \citet{Lis11b} M-R implicitly incorporates compositional variation from planet to planet, it does not allow for the possibility of a multi-valued M-R.  This is potentially a severe shortcoming, as a more complex M-R has appeared to be necessary from the outset of observational constraints on low-mass planet compositions: the first two planets with measured radii and masses, CoRoT-7 b \citep{Que09, Leg09} and GJ 1214 b \citep{Char09}, yielded very different densities (6 g cm$^{-3}$ and 2 g cm$^{-3}$ respectively), despite having similar masses ($4.9$ M$_\oplus$ and $6.5$ M$_\oplus$).  

With this observational evidence in mind, we believe that the key to reconciling the \emph{Kepler} and HARPS occurrence rates may be a multi-valued low-mass M-R.  Our parameterization assumes that the simulated planets can have either a significant gaseous composition (Neptune analogs; an extension of the gas giants to lower masses) or a rocky composition (Earth analogs; an extension of the terrestrial planets to higher masses), and that the admixture of these two compositions varies as a linear function of mass for $1 \leq M \leq 17$ M$_\oplus$.  This admixture is quantified by the fraction of rocky planets in the population, $f_{rocky}(M)$: if, for example, $f_{rocky}(1) = 1.0$ and $f_{rocky}(17) = 0.0$, then $f_{rocky}(8) = 0.5$, meaning that all $1$ M$_\oplus$ planets would be rocky, all $17$ M$_\oplus$ planets would be gaseous, and the $8$ M$_\oplus$ planets would be evenly divided between the two compositions.  In our simulations we allow $f_{rocky}(1)$ and $f_{rocky}(17)$ to vary between 0 and 1 in increments of 0.1, giving two more free parameters in our simulations.  

For the Earth analogs in this multi-valued M-R we use the Solar System's terrestrial planet population-wide mass-to-radius relationship: $R/R_\oplus = (M/M_\oplus)^{0.33}$.  We emphasize that this rocky M-R is not just a re-expression of the individual mass-to-radius relationship for a constant-density sphere; instead, this population-wide M-R was derived by fitting a power law to all of the Solar System's inner planets, much like the $R \propto M^{0.48}$ relationship was derived above.

For the Neptune analogs we use the M-R curves calculated by \citet{Rog11}.  These authors model the structure of low-mass planets with substantial gaseous envelopes by invoking a core accretion formation history and then self-consistently incorporating the effect of planetary equilibrium temperature, $T_{eq}$, across the range of orbital periods and stellar fluxes that we consider here.  They find, however, that the M-R curves of constant gaseous envelope mass fraction, $M_{env}$, are remarkably insensitive to planet mass above $\sim 7$ M$_\oplus$.  Because no single $M_{env}$ provides the dynamic range needed to explain the diversity of radii that \emph{Kepler} observes, we must allow for variation in envelope fraction to construct a population-wide M-R that can reasonably reproduce the observed radius range.  Noting that the M-R curves are roughly equally spaced in $R$ by approximately logarithmic bins in $M_{env}$, we randomly choose an envelope mass fraction from a log-uniform distribution between $10^{-5}$ and $10^{-1}$.  Finally, using Figure 4 of \citet{Rog11}, we interpolate our simulated planets' radii as a function of $M$, $M_{env}$, and $T_{eq}$.

As \citet{Rog11} illustrates, varying $M_{env}$ allows planets with masses as small as $2$ M$_\oplus$ to have a radius as large as $7$ R$_\oplus$, which enables planets less massive than Neptune to fall within the $2 \leq R \leq 6$ R$_\oplus$ range that \emph{Kepler} has found to be well populated \citep{Bor11, How11}.  However, these relatively low-mass, large-radius planets are particularly susceptible to atmospheric mass loss, and so these planets may not actually be able to hold onto their gaseous envelopes, depending on the amount of irradiation they receive from their host star.  Following the discussion in \citet{Rog11}, we incorporate the possibility of mass loss in our population-wide M-R via the following timescale argument.

As illustrated by \citet{Lam03}, one must consider the effect that X-ray and extreme ultraviolet (XUV) irradiation has on a planet's thermal structure in order to realistically treat atmospheric mass loss.  In the regime where the amount of energy incident on the planet determines the degree of atmospheric escape, this mass loss is parameterized by \citep{Lec07, Val10, Rog11}
\begin{equation} \label{massloss}
\dot{M} = - \frac{\epsilon \pi F_{XUV} R^2_{XUV} R_p}{G M_p K_{tide}},
\end{equation}
where $F_{XUV}$ is the XUV flux incident on the planet from the host star; $\epsilon$ is the fraction of incident XUV energy that is actually absorbed by the atmospheric particles; $R_{XUV}$ is the planet radius at which the XUV flux is absorbed; $R_p$ is the radius of the planet as calculated from planetary interior structure models; $M_p$ is the mass of the planet; and $K_{tide}$ is a tidal correction factor of order unity for planets with $R \lesssim R_{Nep}$ and $P > 2$ days.  Unfortunately, $\epsilon$ is largely unknown, so at best Equation \ref{massloss} provides an order-of-magnitude estimate for $\dot{M}$.  We follow \citet{Rog11} in setting $\epsilon = 0.1$ and $F_{XUV} = F_{XUV,\odot} = 4.6 \times 10^{-3}$ W m$^{-2}$ \citep{Rib05}; we scale $F_{XUV}$ by the equilibrium temperature of the planet, which depends on the radius of the host star, the effective temperature of the host star, and the semimajor axis of the planet's orbit.  From the mass loss timescales plotted by \citet{Rog11} we estimate that $R^2_{XUV} \sim 10 R_p^2$ for these short-period low-mass planets.  With $\dot{M}$ thus determined, the atmospheric mass loss timescale is
\begin{equation}
t_{loss} = - \frac{M_{envelope}}{\dot{M}}.
\end{equation}
If $t_{loss} < 1$ Gyr, we consider the planet to have completely lost its gaseous envelope, and we take the radius of the planet to be the radius of its 50\% rock, 50\% ice core \citep{For07}.

\subsection{Star Selection} \label{Selstar}

Once we apply an M-R to the simulated RV population, we randomly allocate planets to specific \emph{Kepler} target stars.  This one-to-one matching allows us to sidestep the concern that the selection biases exhibited by different detection methods will significantly influence computed planet occurrence rates \citep{How11}, and we can directly compare our simulated population with \emph{Kepler}'s planet candidates.  Accordingly, we adopt the list of 165,000 long-cadence Q2 \emph{Kepler} target stars to initiate our star selection.  We begin by extracting the photometrically-derived effective temperature, $T_{eff}$, the surface gravity, log($g$), the radius, $R_{star}$, and the \emph{Kepler}-bandpass apparent magnitude, $Kp$, from the each star's Q2 FITS header.  These data originate from the Kepler Input Catalog (KIC; \citealp{KIC09}), which has known errors of $\pm 200$ K on $T_{eff}$ and $\pm 0.4$ dex on log($g$) \citep{Bro11}.  Because these two parameters are used to calculate $R_{star}$, the errors on the planet candidates' radii can be significant; in \S \ref{Discuss} we discuss the possible effect of these errors on our results.

\begin{figure*}[t]
\epsscale{1.2}
\plotone{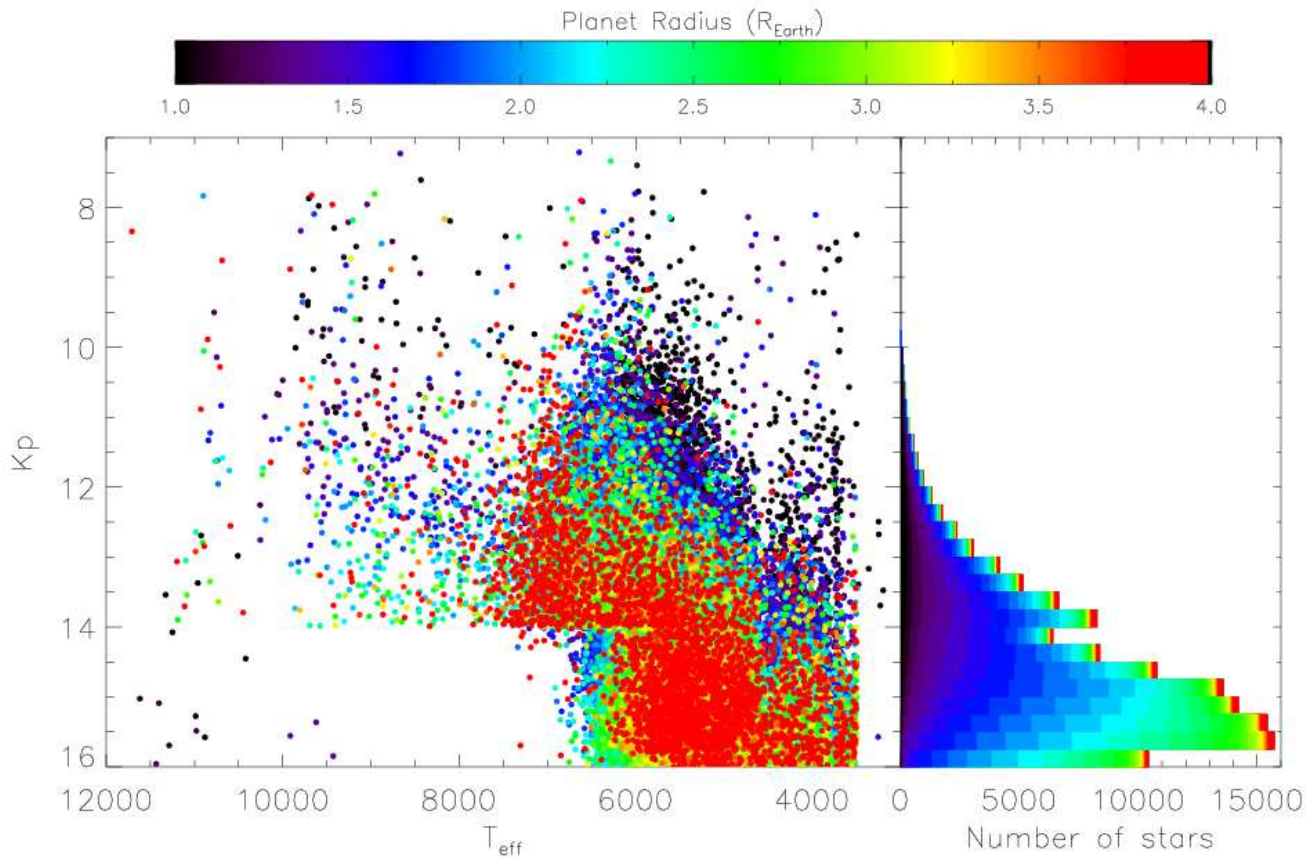}
\caption{Right: apparent magnitude, $Kp$, and effective temperature, $T_{eff}$, from the Kepler Input Catalog (KIC) for the \emph{Kepler} target stars included in our simulations (\S \ref{Selstar}).  All of these stars have KIC log($g$) $> 4.0$.  Left: number of log($g$) $> 4.0$ \emph{Kepler} target stars in each apparent magnitude bin.  The color represents the smallest planet around each target star that \emph{Kepler} could have detected in its first four months of data, assuming an orbit with $P = 20$ days and $e = 0$.  With the same orbital parameters for each size planet, this minimum detectable radius is thus determined by the radius of the star, $R_{star}$, and by the star's total photometric noise on a three-hour timescale, $CDPP_3$ (\S \ref{Detect}). In the scatterplot, note the general trend of minimum detectable radius with both $Kp$ and $T_{eff}$, which correlate with $CDPP_3$ and $R_{star}$, respectively.  The histogram to the right more clearly illustrates the trend of increasing minimum detectable radius with increasing $Kp$ (due to increasing $CDPP_3$).  However, it is important to note that there do exist dim target stars around which \emph{Kepler} could have already detected a 1 to 1.5 $R_\oplus$ planet.  This is a result of the trend of decreasing minimum radius with decreasing $R_{star}$ and the fact that low-mass stars exist in every $Kp$ bin. } \label{figstars}
\end{figure*}

In their analysis of \emph{Kepler}'s planet candidates, \citet{How11} compute \emph{Kepler}'s observed occurrence rates from a heavily vetted list of target stars, whose total noise in one quarter of data enables detection of a $R \geq 2$ R$_\oplus$ planet with SNR $\geq 10$.  This approach prompts them to drop all stars with $Kp \geq 15$ and all planets with $R < 2$ R$_\oplus$ due to concerns about sample incompleteness.  By contrast, our approach retains the entire \emph{Kepler} target star sample, with only the log($g$) cut discussed below: because we individually simulate each planet's light curve to accurately determine its detectability (\S \ref{Detect}) and then ask how many planets \emph{Kepler} would have seen in its first four months of data if the HARPS occurrence rate is true (\S \ref{Results}), we naturally account for the incompleteness in \emph{Kepler}'s first four months of data. (This incompleteness is displayed graphically in Figure \ref{figstars} as the smallest-radius planet that each star could have detected by the end of Q2.) Thus, our simulation procedure permits us to include dimmer stars and smaller planets with radii down to $1$ R$_\oplus$, which allows us to draw conclusions with a larger sample size and improved statistics.

The only severe cut we make to the 165,000 available Q2 target stars is in log($g$).  We restrict potential planet-hosting stars to those with log($g$) $> 4.0$ to minimize contamination from subgiants, as the KIC's surface gravities are poorly contrained above $T_{eff} \sim 5400$ K \citep{Bro11}.  The resulting list consists of 131,000 stars (Figure \ref{figstars}), the vast majority ($> 110,000$) of which are G and K dwarfs.  Nonetheless, a small proportion of subgiants and giants, whose radii may be underestimated in the KIC by as much as a factor of 2 \citep{Bro11}, likely remains in our target star sample.  Without knowledge about the degree of subgiant contamination, we cannot accurately account for their statistical effect in our results, although we expect that this effect will be very small based on the low numbers of possibly misclassified evolved stars found by \citet{Bas11}.

\subsection{Detectability of Simulated Planets} \label{Detect}

To pinpoint the simulated planets that \emph{Kepler} would have identified as planet candidates after four months of data collection, we first compute analytic light curves \citep{Man02} for the simulated planets that transit according to their geometric transit probability \citep{Sea03}.  These light curves incorporate the planets' eccentricity and inclination as well as the \emph{Kepler}-bandpass limb darkening coefficients that are calculated by \citet{Cla11} for a large range of stellar effective temperatures, surface gravities, and metallicities.  Using a 30-minute cadence over 132 days to match \emph{Kepler}'s long-cadence Q0 - Q2 datasets, we determine the transit depth, duration, and the total number of transit events directly from the simulated light curves.

As described in \citet{Bat10}, \emph{Kepler}'s detectability criterion is set such that $< 1$ false positive planet detection over its 3.5 year mission would result from purely statistical fluctuations in stellar photon counts.  This requirement gives a $7.1 \sigma$ threshold for a transit's statistical significance when the light curve is folded and binned.  The detectability of a planet therefore depends on both $R_{pl}/R_{star}$ and a number of stellar parameters and instrumental properties which affect the total noise \citep{Bat10, Jen10}.  These systematic errors are difficult to assess without intimate knowledge of \emph{Kepler}'s performance, so we use the noise calculated directly by the \emph{Kepler} data reduction pipeline, the Combined Differential Photometric Precision (CDPP, obtained from J. Christiansen \& J. Jenkins via personal communication, June 7, 2011), to reproduce as accurately as possible the planet population that \emph{Kepler} could have identified by the end of Q2.

Defined as the root mean square of stellar photometric noise on transit timescales, the CDPP provides the most accurate estimate of the noise from each target star that would interfere with a transiting planet's detectability.  A wavelet-based, adaptive matched filter is applied to the corrected \emph{Kepler} light curves in the Transiting Planet Search section of the Science Processing Pipeline \citep{Jen10} to produce 3-hour, 6-hour, and 12-hour CDPP estimates, which are then used to calculate the statistical significance of a possible transit event.  Incorporating \emph{Kepler}'s own noise metric in our simulations automatically folds in its detection biases and accounts for sample incompleteness below $2$ R$_\oplus$; therefore, we can extend our analysis down to Earth-sized planets without reservations about hidden selection effects.  

Our simulations only consider planets with $2 \leq P \leq 50$ days, so the 3-hour CDPP estimate is the most relevant for our purposes.  Matching each planet to a \emph{Kepler} target star also matches it to a CDPP value, so we scale this noise estimate by the transit duration and the total number of transit events observed during Q0 - Q2 \citep{Bat10, How11}.  Our detectability criterion therefore becomes:
\begin{equation} \label{detecteq}
SNR =  \frac{\delta \sqrt{N_{tr}\frac{ N_{dur}}{6}}}{CDPP_3} > 7.1,
\end{equation}
where $\delta \propto (R_{pl}/R_{star})^2$ is the maximum transit depth (in ppm) identified from the analytic light curves, $CDPP_3$ is the Q2 3-hour Combined Differential Photometric Precision (in ppm) associated with the planet's host star, $N_{tr}$ is the number of observed transits in four months, and $N_{dur}$ is the number of data points acquired per transit on a 30-minute cadence.  We note that $\delta$ is proportional but not equal to $(R_{pl}/R_{star})^2$ because we include a range of possible transit-producing inclinations and self-consistently incorporate the effect of limb darkening based on the host star's $T_{eff}$ and log($g$).

Figure \ref{figstars} illustrates our detectability criterion graphically, with the color scale showing the smallest planet for each log($g$) $> 4.0$ target star that \emph{Kepler} could have detected after four months of data collection, assuming an orbit with $P = 20$ days and $e = 0$.  As expected, this minimum detectable $R_{pl}$ trends with both $Kp$ and $T_{eff}$, which correlate with CDPP and $R_{star}$, respectively.  When the orbit is not held constant, an individual planet's detectability is also determined by its orbital period, as given by $N_{tr}$ in Equation \ref{detecteq}.

\subsection{Statistics of the Detectable Period-Radius Distribution} \label{Stats}

\begin{figure}[b]
\epsscale{1.2}
\plotone{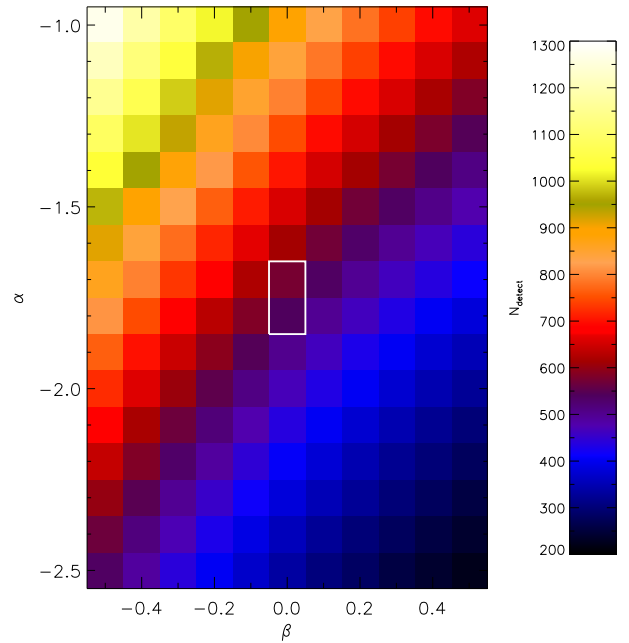}
\caption{The total number of detectable (\S \ref{Detect}) simulated planets ($N_{detect}$) produced by the $R = M^{0.48}$ mass-to-radius relationship (\S \ref{singleMtoR}) for a 40\% overall occurrence rate in the 2 $\leq P \leq 50$ days and $1 \leq R < 4$ R$_\oplus$ range; this total number is averaged over 100 realizations.  As a point of reference, the total number of \emph{Kepler} planet candidates that we use for this comparison is 631.  The axes denote the period power law index ($\beta$) and mass power law index ($\alpha$), which serve as free parameters in our simulations (\S \ref{Define}).  The white box outlines the values for $\alpha$ and $\beta$ which produce a median probability $\mathcal{P}_{sing} > 10^{-4}$ over all N=100 realizations.} \label{totnum33}
\end{figure}

The above procedure gives us the period-radius distribution that \emph{Kepler} would observe when the underlying planet population conforms to the HARPS occurrence rate.  Our lack of detailed knowledge about the HARPS data set, however, has introduced some freedom in the population's initial mass and period distributions.  Different input distributions can significantly affect the total number of planets that are detectable by \emph{Kepler} (\S \ref{singleres}; Figure \ref{totnum33}), so we must identify the free parameters which produce detectable planet distributions that best fit \emph{Kepler}'s planet candidates before we can conclude that any discrepancy between the total number of observed planets is a result of the overall 30-50\% statistic.  A general sense of the appropriate $\alpha$'s and $\beta$'s can be gleaned from the analyses of \citet{How11} and \citet{You11}, but because we begin with a Doppler survey rather than a transit survey and because we consider multiple-valued mass-to-radius relationships, an independent assessment of the best-fit mass and period distributions is valuable.

Before we can make this comparion, however, we need to filter the list of 1,235 \emph{Kepler} planet candidates to match the limits we impose on the simulated population.  Accordingly, we retain only those candidates with $1 \leq R < 4$ R$_\oplus \sim R_{Nep}$ and $2 \leq P \leq 50$ days orbiting stars with $Kp \leq 16$.  We also impose a cut on the candidates in multiple-planet systems, including only the first planet listed by the \emph{Kepler} Science Processing Pipeline in this mass and period range; in most cases, this is the planet labeled ``.01".  This cut conforms to our assumption of one planet per host star and reduces the total number of \emph{Kepler} planets in our radius and period range from 797 to 631, a difference of 166 ($\sim 20\%$).  Depending on the multiple-planet prescription that could be applied to the HARPS occurrence rate, our single-planet assumption may either overestimate or underestimate the difference in the total number of planets between the simulation and the full \emph{Kepler} low-mass data set.

To address the goodness-of-fit between the \emph{Kepler}-detectable, simulated period-radius distribution and the \emph{Kepler} candidates' period-radius distribution, we employ the two-sample two-dimensional Kolmogorov-Smirnov (2-D K-S) test \citep{Fas87} as well as the sample size-independent one-dimensional $\chi^2$ test over planet radius.  The 2-D K-S test avoids binning data, unlike the $\chi^2$ test, and thus maximally preserves information contained in the radius-period distribution of the detectable simulated planets.  However, the K-S test is most sensitive to the middle of its data range, and so we use the 1-D $\chi^2$ test to distinguish between a 2-D K-S statistic produced by a good fit in the $2$ R$_\oplus \lesssim R \lesssim 3$ R$_\oplus$ range and a similar 2-D K-S statistic produced by a good fit over the entire $1$ R$_\oplus \leq R < 4$ R$_\oplus$ range.  We use the sample size-independent version of the $\chi^2$ statistic to fairly assess the goodness-of-fit between different power-law indeces, as this fit would otherwise be dominated by differences in the total number of detectable planets.  

In practice, using the $\chi^2$ test for the planet radius distribution does not discard much information, as the \emph{Kepler} planet candidates' radii, rounded to the nearest $0.1$ R$_\oplus$, are already effectively binned.  To make sure that this artificial structure in the \emph{Kepler} period-radius distribution does not determine the quality of the best fit, we also round the simulated planets' radii to the nearest $0.1$ R$_\oplus$ before computing the 2-D K-S statistic.  Thus, the 2-D K-S test's added value is the simultaneous inclusion of the period distribution with the radius distribution for a comprehensive goodness-of-fit.


\section{The Transit-RV Comparison: Results} \label{Results}

\subsection{Single-Valued M-R} \label{singleres}

The result of 100 realizations of the $R/R_\oplus = (M/M_\oplus)^{0.48}$ mass-to-radius relationship (\S \ref{singleMtoR}) computed at a 40\% overall occurrence rate is illustrated in Figure \ref{totnum33}.  The color denotes the total number of detectable (\S \ref{Detect}) simulated planets ($N_{detect}$) with $1 \leq R \leq 3.9$ R$_\oplus = (17$ M$_\oplus)^{0.48}$ and 2 $\leq P \leq 50$ days, averaged over all N=100 realizations; the total number of analogous \emph{Kepler} planet candidates in our filtered list is 631.  Figure \ref{totnum33} indicates that $N_{detect}$ depends sensitively on the mass power law index $\alpha$ and period power law index $\beta$ (\S \ref{Define}).  Thus, we need to identify the $\alpha$ and $\beta$ which give the best fit between the \emph{Kepler}-detectable population and \emph{Kepler}'s planet candidates in order to conclude that any discrepancy between the total number of observed planets is a result of the overall 30-50\% statistic, not a result of using an inappropriate power law.

Calculating both the 2-D K-S test and the 1-D $\chi^2$ test allows us to identify these best-fitting free parameters (\S \ref{Stats}).  The 2-D K-S test gives us the probability $\mathcal{P}$ that the two populations were drawn from the same underlying radius and period distribution; for the single-valued M-R considered here, the maximum $\mathcal{P}_{sing} \sim 10^{-4}$.  In Figure \ref{totnum33} we have boxed in white the values for $\alpha$ and $\beta$ which produce a median $\mathcal{P}_{sing} > 10^{-4}$ over all N=100 realizations: $\alpha = -1.8$, $\beta = 0.0$, and $\alpha = -1.7$, $\beta = 0.0$.  These values correspond to $N_{detect} = 537 \pm 28$ and $N_{detect} = 574 \pm 28$, respectively.  Thus, a 40\% occurrence rate for $R/R_\oplus = (M/M_\oplus)^{0.48}$ actually under-predicts the total number of planets that \emph{Kepler} would see in its first four months of data.

\begin{figure}[t]
\epsscale{0.5}
\includegraphics[angle=90,scale=0.35]{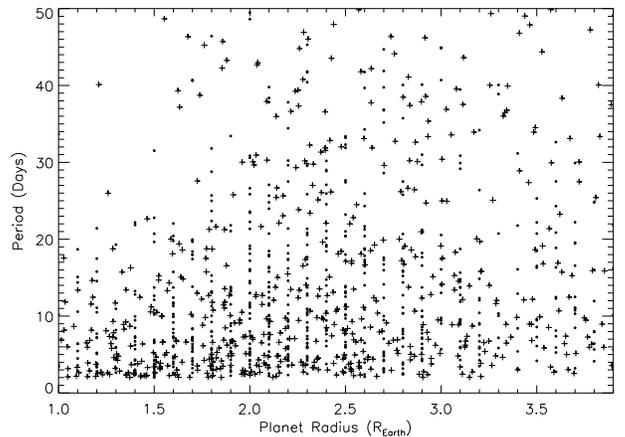}
\caption{Period vs. radius for a single realization of the simulated planet population produced by the $R = M^{0.48}$ mass-to-radius relationship with $\alpha = -1.8$, $\beta = 0.0$, and a 40\% overall occurrence rate.   The \emph{Kepler} planet candidates are marked with the circles, and the detectable simulated planets are marked with the plus signs.  Comparing the two data sets yields a 2-D K-S probability $\mathcal{P}_{sing} = 0.03\%$.  Note the relative paucity of simulated planets in the $1.8 \lesssim R \lesssim 3$ R$_\oplus$ and $P < 20$ days range.} \label{perrad33}
\end{figure}

\begin{figure}[b]
\epsscale{0.5}
\includegraphics[angle=90,scale=0.35]{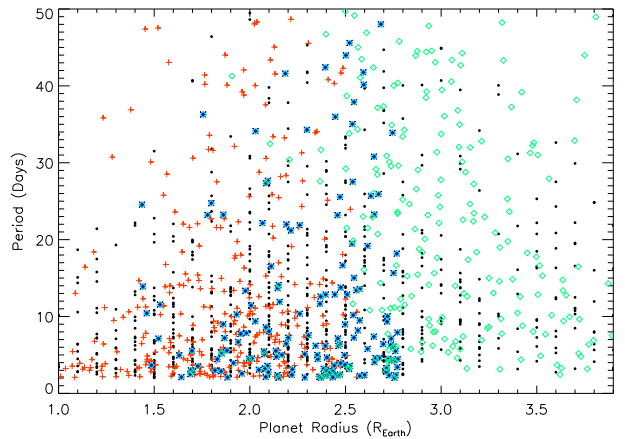}
\caption{Period vs. radius for a single realization of the simulated planet population produced by the multi-valued mass-to-radius relationship with $\alpha = -1.0$, $\beta = 0.0$, $f_{rocky}(1) = 0.9$, $f_{rocky}(17) = 0.1$, and a 40\% overall occurrence rate.   The black circles denote the \emph{Kepler} planet candidates; the red plus signs denote the detectable simulated planets with a rocky composition; the green diamonds denote the detectable simulated planets with a gaseous composition; and the blue asterisks denote the detectable simulated planets with a half-rock, half-ice composition, which could be produced by significant mass loss from the gaseous planets.    Comparing the two data sets yields a 2-D K-S probability $\mathcal{P}_{mult} = 0.13\%$.} \label{perrad32}
\end{figure}

Before we can discuss the implications of this result, however, we must address the low probabilities produced by the 2-D K-S test.  Figure \ref{perrad33}, representing one realization of the data at $\alpha = -1.8$ and $\beta = 0.0$, sheds some light on the cause of this issue: this single-valued M-R does not produce enough planets with $1.8 \lesssim R \lesssim 3$ R$_\oplus$ and $P < 20$ days.  Our treatment of \emph{Kepler}'s sample incompleteness does not seem to be at fault, as there are very few detectable simulated planets at small radii and long periods where \emph{Kepler}'s detectability criterion (\S \ref{Detect}) rejects the most planets.  Furthermore, our use of \emph{Kepler} target stars (\S \ref{Selstar}) precludes \emph{Kepler}'s selection biases as an explanation for this discrepancy.  We therefore turn to a more flexible mass-to-radius relationship as a potential means to improve the transit-RV fit.

\subsection{Multi-Valued M-R} \label{multires}

\begin{figure}[b]
\epsscale{1.2}
\plotone{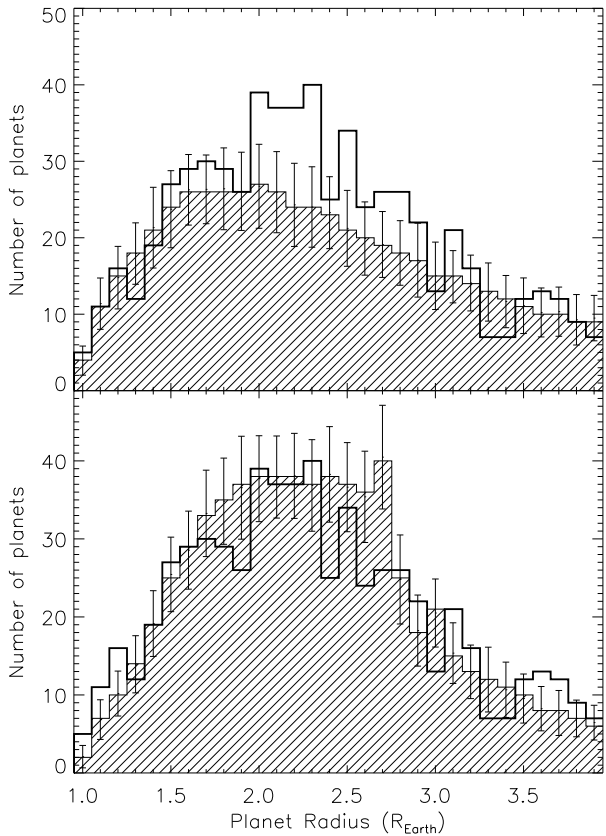}
\caption{Number of planets vs. radius averaged over all N=100 realizations of the simulated planet population (striped pattern); the error bars denote the standard deviation of the number of detectable planets in each bin. The radius distribution of the \emph{Kepler} planet candidates is displayed as the thick black line.  Top: the single-valued M-R with $\alpha = -1.8$, $\beta = 0.0$, and a 40\% overall occurrence rate, for which $\chi^2 = 1.2$.  Bottom: the multi-valued M-R with $\alpha = -1.0$, $\beta = 0.0$, $f_{rocky}(1) = 0.9$, $f_{rocky}(17) = 0.1$, and a 40\% overall occurrence rate, for which $\chi^2 = 1.8$.  This $\chi^2$ value is slighly higher due to the smaller error bars in the wings where the two data sets also happen to produce a poorer match.} \label{hist}
\end{figure}

Figure \ref{perrad32} displays the period-radius distribution for one realization of our multi-valued M-R (\S \ref{multiMtoR}).  A qualitative comparison with Figure \ref{perrad33} indicates that the multi-valued M-R does give a better fit than the single-valued M-R, which is corroborated quantitatively by an order-of-magnitude improvement in the 2-D K-S probability ($\mathcal{P}_{mult} \sim 0.1\%$).  Unfortunately, however, this probability is still very low.  To isolate the discrepancy, we compute $\chi^2$ for the radius distribution averaged over 100 realizations of the simulated planet population; these distributions are shown in Figure \ref{hist} and correspond to $\chi^2 = $ 1.2 and 1.8.  The fact that the reduced $\chi^2$ values are close to unity is encouraging and indicates that the discrepancy lies in the simulated planets' period distribution, which the 2-D K-S test has the leverage to assess (\S \ref{Stats}).  

A discrepancy in period is not unexpected considering the simplifying assumptions we made to the input period distribution (\S \ref{Define}).  Given that $\mathcal{P}_{mult} > \mathcal{P}_{sing}$, however, the detectable period distribution also depends on the population-wide M-R.  It is probable that other population-wide M-Rs could further improve the transit-RV fit, but it remains unclear whether these M-Rs would be parameterizable in a reasonable number of degrees of freedom.  We therefore chose simplicity over absolute best fits to offer M-Rs that are both physically intuitive and computationally feasible.

\begin{figure*}[t]
\epsscale{1.1}
\plotone{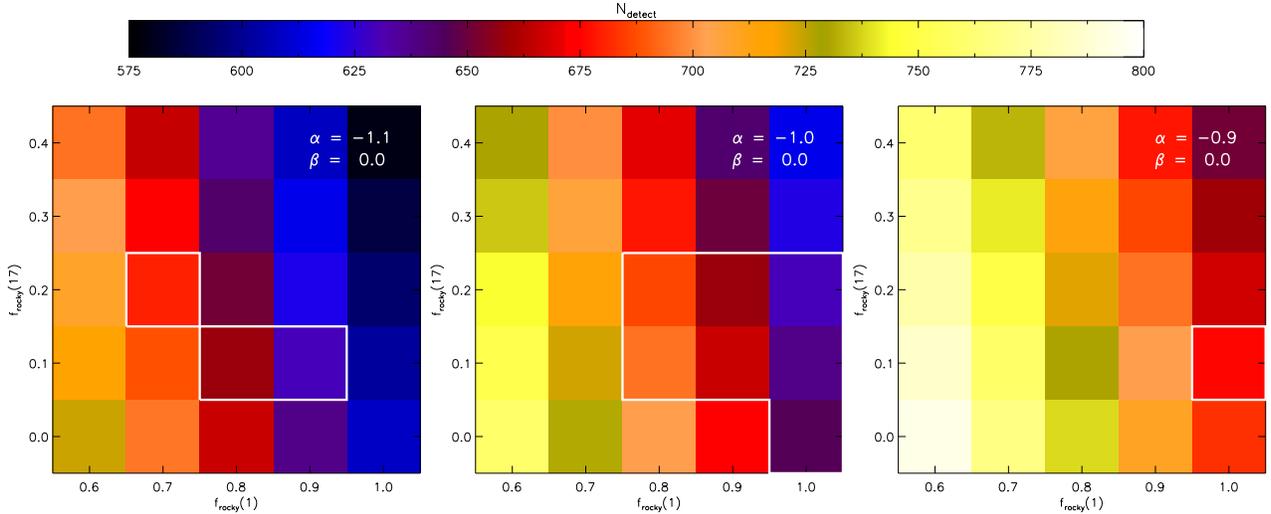}
\caption{The total number of detectable (\S \ref{Detect}) simulated planets ($N_{detect}$) with 2 $\leq P \leq 50$ days and $1 \leq R < 4$ R$_\oplus$ produced by the multi-valued mass-to-radius relationship (\S \ref{multiMtoR}) for a 40\% overall occurrence rate; this total number is averaged over 100 realizations.  As a point of reference, the total number of \emph{Kepler} planet candidates that we use for this comparison is 631.  The axes denote the fraction of all $1$ M$_\oplus$ planets in the simulated planet population that have a rocky composition, $f_{rocky}(1)$, and the fraction of all $17$ M$_\oplus$ planets that have a rocky composition, $f_{rocky}(17)$; each panel corresponds to a different value of the mass power law index ($\alpha$) at a constant period power law index of $\beta = 0.0$.  The white boxes outline the values for $f_{rocky}(1)$ and $f_{rocky}(17)$ which produce both a median probability $\mathcal{P}_{mult} > 0.1\%$ over all N=100 realizations and a reduced $\chi^2 \leq 2.0$ for the radius distributions analogous to those in Figure \ref{hist}.} \label{totnum32}
\end{figure*}

Figure \ref{totnum32} shows the total number of detectable simulated planets ($N_{detect}$) averaged over 100 realizations of the simulated population for a 40\% overall occurrence rate.  The axes correspond to the two free parameters that characterize the multi-valued M-R in addition to $\alpha$ and $\beta$: (1) the fraction of all $1$ M$_\oplus$ planets that have a rocky composition, $f_{rocky}(1)$, and (2) the fraction of all $17$ M$_\oplus$ planets with a rocky composition, $f_{rocky}(17)$.  As with Figure \ref{totnum33}, the best fitting free parameters in Figure \ref{totnum32} are outlined with white boxes; in this case, a good fit is identified by a median $\mathcal{P}_{mult} > 0.1\%$ over all 100 realizations of the simulated population and a reduced $\chi^2 \leq 2.0$ for the radius distributions analogous to those in Figure \ref{hist}.  The values of these statistics as well as the total numbers of detectable simulated planets are listed in Table \ref{bestfits32}. 

\begin{deluxetable}{ccccccc}
\tablecaption{Total Number of Detectable Planets for the Best-fitting Parameters of the Multi-Valued M-R and a 40\% HARPS Occurrence Rate\label{bestfits32}} 
\tablewidth{0pt}
\tablehead{
\colhead{$f_{rocky}(1)$} & \colhead{$f_{rocky}(17)$} & \colhead{$\alpha$} &
\colhead{$\beta$} & \colhead{$\mathcal{P}_{mult}$} & \colhead{$\chi^2$} & \colhead{$N_{detect} (40\%)$} 
}
\startdata
\\
0.7 &	 0.2 & -1.1 & 0.0 & 0.11\% & 2.0 & 679 $\pm$ 27 \\[2pt]
0.8 &	 0.1 & -1.1 & 0.0 & 0.11\% & 1.7 & 657 $\pm$ 26 \\[2pt]
0.8 &	 0.1 & -1.0 & 0.0 & 0.11\% & 1.9 & 694 $\pm$ 27 \\[2pt]
0.8 &	 0.2 & -1.0 & 0.0 & 0.16\% & 2.0 & 685 $\pm$ 27 \\[2pt]
0.9 &	 0.1 & -1.1 & 0.0 & 0.11\% & 1.6 & 629 $\pm$ 27 \\[2pt]
0.9 &	 0.1 & -1.0 & 0.0 & 0.18\% & 1.8 & 665 $\pm$ 28 \\[2pt]
0.9 &	 0.2 & -1.0 & 0.0 & 0.17\% & 1.9 & 657 $\pm$ 28 \\[2pt]
1.0 &	 0.0 & -1.0 & 0.0 & 0.14\% & 1.6 & 645 $\pm$ 27 \\[2pt]
1.0 &	 0.1 & -1.0 & 0.0 & 0.18\% & 1.7 & 637 $\pm$ 27 \\[2pt]
1.0 &	 0.1 & -0.9 & 0.0 & 0.21\% & 2.0 & 675 $\pm$ 29 \\[2pt]
1.0 &	 0.2 & -1.0 & 0.0 & 0.12\% & 2.0 & 629 $\pm$ 27 \\
\enddata
\end{deluxetable}

$N_{detect}$ varies with $f_{rocky}(1)$ and $f_{rocky}(17)$ and depends strongly on $\alpha$ and $\beta$.  Given that the total number of \emph{Kepler} planet candidates in our filtered list is 631, the best-fit free parameters yield total numbers of detectable planets that are similar between the two data sets.  Before any conclusions can be drawn from this, however, we must account for the probability of false positives among \emph{Kepler}'s planet candidates.  Fortunately, the 5 - 15\% false alarm rate, which could lower the total number of actual planets in our filtered list to 535 in the worst-case scenario, falls within HARPS' $\pm 10\%$ error bars: because $N_{detect}$ is directly proportional to the overall occurrence rate, \emph{Kepler} would have detected $\sim 500$ planets in its first four months of data if 30\% of main sequence stars host at least one planet, with $\alpha = -1.0$, $\beta = 0.0$, $f_{rocky}(1) = 0.9$, and $f_{rocky}(17) = 0.1$ parameterizing the population.  Thus, our multi-valued M-R gives total numbers of detectable simulated planets that are similar to \emph{Kepler}'s total number of planet candidates even when the false alarm probability is taken into account, indicating that the HARPS overall 30-50\% statistic \emph{can in fact be consistent} with \emph{Kepler}'s results.

\section{Discussion} \label{Discuss}

Because population-wide mass-to-radius relationships (M-R) are central to transit-RV comparison studies, they require realistic treatments of planet composition across the entire population.  Ideally the requisite assumptions about the population's bulk density distribution would be informed by observations of planets that are detected by both methods.  Unfortunately, however, current observations of sub-Neptune-mass planets lack sufficient data to address this issue: Kepler-11 b - f \citep{Lis11a}, whose mass measurements have significant errors ($\pm$ 30 - 100\%), are the only confirmed transiting planets that fall securely in the mass and period ranges considered in this paper ($1 \leq M \leq 17$ M$_\oplus$ and $2 \leq P \leq 50$ days).  A single-valued mass-to-radius relationship derived these few dually-detected low-mass exoplanets (\S \ref{singleMtoR}) furthermore provides a poor fit the \emph{Kepler} planet candidates' period-radius distribution (\S \ref{singleres}), suggesting that the assumptions such a power law makes about the population's density distributions are incorrect.

Given that the first two dually-detected ``Super-Earth" planets have similar masses but different bulk densities (CoRoT-7 b: \citealp{Que09}, \citealp{Leg09}; GJ 1214 b: \citealp{Char09}), it is not surprising that a single-valued M-R produces poor agreement between transit and RV survey results.  Hints of a multi-valued M-R that allows planets with different densities to occur at the same mass have continued to emerge with more recent detections: most of the Kepler-11 planets have low bulk densities (0.5 - 3.1 g/cm$^3$; \citealp{Lis11a}), while Kepler-10 b and 55 Cnc e yield densities of 9 g/cm$^3$ \citep{Bat11} and 5 - 6 g/cm$^3$ \citep{Win11, Dem11}, respectively.  A popular explanation for this compositional bimodality is that the high-density planets, which so far are all observed on extremely close-in orbits ($P < 2$ days), constitute the special case of low-mass gas planets that have had their atmospheres completely stripped, leaving only their solid cores behind \citep{Scha09, Jac10, Bat11}.  Instead, we propose that these high-density planets constitute a more general short-period --- and thus more easily detectable --- case of an entirely different class of exoplanets: true super-Earths that formed with a primarily refractory composition.  This new interpretation has significant implications for planet formation (i.e. \citealp{Han11}), suggesting that there may be multiple modes of formation for planets in this mass range \citep{Leg11}.
To ascertain whether this is the case, however, we must break the degeneracy between the two interpretations.  The current state of observational data on individual planets cannot accomplish this task, 
but a statistical transit-RV study such as the one conducted here can potentially elucidate the correct interpretation: considering transit surveys' bias toward larger and thus lower density planets, and RV surveys' bias towards more massive and thus higher density planets, statistical discrepancies between the observed radius and period distributions could indicate that a complex population-wide M-R is at play.

As a result, we believe that a multi-valued M-R is crucial for both explaining the apparent discrepancy between the \emph{Kepler} and HARPS occurrence rates and for developing a full understanding of the Galactic planetary population.  The multi-valued M-R we present here (\S \ref{multiMtoR}) adopts two compositions: rocky planets that follow the same $R/R_\oplus = (M/M_\oplus)^{0.33}$ relationship as the Solar System's inner planets, and gaseous planets that follow the M-R curves presented in \citet{Rog11}, while a prescription for atmospheric mass loss introduces a third intermediate composition.  An admixture of these compositions over the entire 1 M$_\oplus \leq M \leq 17$ M$_\oplus$ mass range is able to account for the density variation that exists among low-mass planets.  We emphasize that the order-of-magnitude mass loss prescription we appeal to here does not attempt to model the details of atmospheric escape; we use it only as a way to account for the evolution of a gaseous planet's radius in the low mass, large radius regime.  Interestingly, the presence of intermediate-density planets in a period-radius parameter space unoccupied by rocky or gaseous planets suggests that an intermediate-density planet population, however its constituent planets were formed, is another key component of the transit-RV comparison.

For our multi-valued M-R we have placed particular emphasis on parameterizing the relative contributions from the rocky and gaseous compositions in as physically intuitive a way as possible, while taking care to minimize the number of free parameters.  As a result, we adopt a parameterization that flows naturally from the coexistence of rocky super-Earths and gaseous sub-Neptunes at each planet mass and involves only two additional degrees of freedom: (1) the fraction of all $1$ M$_\oplus$ planets in the simulated planet population that have a rocky composition, $f_{rocky}(1)$, and (2) the fraction of all $17$ M$_\oplus$ planets that have a gaseous composition, $1 - f_{rocky}(17)$, with $f_{rocky}$ varying linearly between the bounding masses.  

Not only is a multi-valued M-R physically intuitive, but it also better fits \emph{Kepler}'s second-quarter planet candidates (\S \ref{multires}).  This improved fit enables us to address the question of planet occurrence rates, as the total number of simulated planets that pass \emph{Kepler}'s detectability criterion (\S \ref{Detect}) can only be attributable to the overall occurrence rate once the two distributions of detectable/detected planets are consistent with each other (\S \ref{singleres}).  As a result, we find that HARPS' 40\% occurrence rate \emph{is in fact consistent} with \emph{Kepler}'s planet candidates for the range of best-fitting parameters in our simulations: $\alpha = -1.1$ to $-0.9$, $\beta = 0.0$, $f_{rocky}(1) = 0.7$ to 1.0, and $f_{rocky}(17) = 0.0$ to 0.2.  The apparent discrepancy between the HARPS and \emph{Kepler} occurrence rates 
 therefore can be naturally explained by the presence of dense planets in the HARPS data set --- planets that, due to their relatively small radii, \emph{Kepler} simply did not find after four months of data collection.

Caution, of course, is in order.  We have made a number of assumptions in our simulations, the most stringent of which was restricting each host star to only one planet (\S \ref{Define}).  We accounted for this by only considering the first planet candidate in our radius and period range to be listed by the \emph{Kepler} pipeline in each multiple-planet system, which generates a $\sim 20\%$ overall reduction of included planet candidates (\S \ref{Stats}).  To be sure, a multiple-planet prescription could be applied to the simulated planets, which would allow all of the \emph{Kepler} planet candidates with $1 \leq R < 4$ R$_\oplus$ and $2 \leq P \leq 50$ days to be included in the comparison.  However, the HARPS occurrence rate offers no information about the appropriate multiple-planet assumptions to make, and including such assumptions only muddies the clear implication of the seemingly discrepant RV and transit occurrence rates.

Nonetheless, our use of only single-planet systems necessitates close consideration of the meaning of an ``occurrence rate" as well as how these ``occurrence rates" are calculated from study to study, as pointed out by both \citet{How11} and \citet{You11}.  The HARPS overall occurrence rate, which makes a statement about the fraction of stars with planets, treats the presence of planets around stars as a binary state: either the star hosts no planets, or it hosts one or more planets, making our single-planet assumption a very natural one to adopt.  On the other hand, the occurrence rates computed by \citet{How11} and \citet{You11} include the possibility of multiple-planet systems and give the number of planets per star (NPPS), rather than the fraction of stars with planets (FSWP).  With information about the distribution of multiple-planet systems such as that offered by \citet{Lat11} and \citet{Tre11}, an NPPS occurrence rate can be directly compared to a FSWP occurrence rate.  For our purposes we simply note that the occurrence rates which \citet{How11} and \citet{You11} compute (0.13 and 0.19 planets per star, respectively, for $2 \leq R < 4$ R$_\oplus$ and $P < 50$ days) would become even lower when transformed to a FSWP occurrence rate, given the presence of multiple-planet systems; this only worsens the apparent discrepancy between the two surveys' occurrence rates.  \citet{You11} does point out, however, that if planets down to $0.5$ R$_\oplus$ are included, then this number-of-planets-per-star occurrence rate may be as high as 1.36.  Thus, for the full $1 \leq R < 4$ R$_\oplus$ range we consider in this paper, the apparent occurrence rate discrepancy may also be explained at least in part by the slight differences in the considered radius range.

Another potential source of concern is the difference between each survey's target star selection criteria.  We address the biases produced from \emph{Kepler}'s selection criteria by drawing from the Q2 targets stars, and we account for its incompleteness by including the Q2 3-hour CDPP measurements; these considerations stem from how we frame the transit-RV comparison, as we ask how many short-period, low-mass planets \emph{Kepler} would have detected in its first four months of data if the HARPS occurrence rate is true.  However, to make a thorough comparison one also needs to consider how these biases differ from the RV selection criteria that factor into HARPS' overall occurrence rate.  Both HARPS and \emph{Kepler} preferentially choose G and K dwarfs with high signal-to-noise ratios \citep{May09,Udr00,Bat10}, but HARPS also targets slowly rotating, magnetically quiet stars and includes no known spectroscopic binaries.  Thus, the differences between the two survey's selection criteria lie in the presence of binary stars in the \emph{Kepler} sample and in the distinction between RV stellar jitter and photometric noise.

According to \citet{Bat10}, \emph{Kepler} searches for planets around all of the known eclipsing binaries ($> 600$) in its field of view.  While these eclipsing binaries are not numerous enough by themselves to appreciably affect our statistics, the unidentified spectroscopic binaries in \emph{Kepler}'s field of view potentially are, if one reasonably allows for the possibility that the planet occurrence rate can differ between single stars and binary systems.  To get a sense for the magnitude of this effect, we refer to \citet{Duq91}, who estimate that as many as two thirds of all G dwarfs have a stellar companion.  The lognormal period distribution they find for spectroscopic G-dwarf binaries indicates that roughly 8\% of all G dwarfs exist in binaries separated by $< 0.5$ AU and $\sim 20\%$ in binaries separated by $\lesssim 10$ AU; considering that \emph{Kepler}'s false-positive vetting process enables binaries at separations of $< 1^{\prime\prime}$ \citep{Bat11} to be identified, the relative fraction of tight binaries in the \emph{Kepler} target star list could be even higher.  Separations of $< 0.5$ AU and $< 10$ AU are especially of interest for the survival and formation of planets in binary systems, as the orbits of the planets considered in this paper would not be stable in equal-mass binary systems separated by $< 0.5$ AU, and protoplanetary disks around the primaries of $\lesssim 10$ AU binary systems would be truncated before the distance at which an ice line could form.  Interestingly, a difference in the planet occurrence rate for binaries with $< 10$ AU separations versus those with $> 10$ AU separations could provide a way to discriminate between the compositions of these close-in planets, assuming that the terrestrial planets formed in-situ and the gaseous planets migrated in from wider orbits.

The HARPS requirement that its target stars have low levels of RV stellar jitter is another potentially significant difference between the two surveys' target selection criteria.  It is certainly the case that \emph{Kepler} has preferentially chosen target stars that exhibit low photometric noise \citep{Bat10}, but this noise is primarily correlated with the apparent magnitude of the star (i.e. Figure \ref{figstars}) and does not necessarily reflect the degree of magnetic activity that heavily factors into the HARPS log($R^\prime_{HK}$) $< -4.8$ target selection.  If we temporarily ignore this, however, and assume that photometric noise is strongly correlated with stellar jitter, we can assess the effect of this selection criterion on our results.  We find that limiting our potential host stars to the $\sim$ 35,000 \emph{Kepler} targets with CDPP$_3 \leq$ 150 ppm worsens the discrepancy between the \emph{Kepler} and HARPS occurrence rates: for $\alpha = -1.0$, $\beta = 0.0$, $f_{rocky}(1) = 0.9$, $f_{rocky}(17) = 0.1$, and a 40\% overall occurrence rate, we find that \emph{Kepler} would have been able to detect 291 $\pm$ 19 planets in its first four months of data (N$_{realizations}=100$), while \emph{Kepler} has actually found 217 planet candidates around stars with CDPP$_3 \leq$ 150 ppm.  A 30\% HARPS occurrence rate is needed to bring these numbers into agreement, making the HARPS-\emph{Kepler} consistency marginal at best, although a high spectroscopic binary fraction in the \emph{Kepler} sample could counteract this effect and improve the consistency between the occurrence rates.  In any case, systematically accounting for the selection of quiet stars requires the forthcoming results of stellar photometric variation studies (i.e. \citealp{Bas11}) to draw conclusions about the \emph{Kepler} target stars' magnetic activity, given the absence of spectra for a majority of these targets.  

In short, we acknowledge that the differences in the two surveys' target star selection criteria could explain some of the apparent discrepancy between their occurrence rates.  Our intent here is simply to point out a plausible, testable explanation for an overall transit-RV occurrence rate discrepancy --- the existence of a distribution of densities in a planet population --- that does not depend on the selection criteria to produce similar numbers of observable planets.

As a final note, we remark that significant errors in the \emph{Kepler} target stars' radii could affect the best-fitting parameters that we find for our multi-valued M-R.  Based on the uncertainty in the Kepler Input Catalog's estimates of $T_{eff}$ and log($g$) \citep{Bro11}, the stellar radii --- and thus the radii of \emph{Kepler}'s planet candidates --- are uncertain by tens of percent.  These errors are substantial; however, the benefit of performing a statistical analysis such as the one presented here is that normally distributed errors will tend to average out, given a large enough sample.  Unfortunately, the errors on the KIC radii are not necessarily normally distributed, and in at least one instance they are known to be severely systematically biased in one direction: the presence of unidentified subgiant stars in the target stars list can underestimate the stellar radii by as much as a factor of 2 \citep{Bro11}.  We have attempted to minimize the effect of such a severe systematic error by limiting the \emph{Kepler} target stars we consider to only those with log($g$) $> 4.0$ (\S \ref{Selstar}), but this does not guarantee that our sample of potential host stars are completely free of systematic biases that could change the best fits we calculate in our simulations.  

Fortunately, in this paper we are more concerned with the total number of detectable planets for its implications about the low-mass planet occurrence rates calculated by different planet detection techniques.  Considering that the $\pm 10\%$ error bars in the HARPS overall occurrence rate can account for the variation in the total number of planets produced by changes in the multi-valued M-R's degrees of freedom (\S \ref{multires}) as well as by a possible 5 - 15\% false positive rate among \emph{Kepler}'s planet candidates \citep{Mor11}, we consider it likely that the HARPS and \emph{Kepler} occurrence rates are actually consistent with each other, with the implication being that there is a distribution of super-Earth/sub-Neptune densities at each planet mass.  We have therefore illustrated the importance of using a multi-valued M-R when comparing RV and transiting planet populations, and we have shown that HARPS is likely detecting a large population of dense low-mass planets.




\acknowledgments

We would like to thank Andrew Howard for a very informative discussion about planetary statistics at the January AAS meeting; Kevin Schlaufman for many helpful conversations about planet formation, the Kepler Input Catalog, and this project in general; Damien S\'egransan and Dan Fabrycky for their feedback on this paper; and Jonathan Fortney for pointers on exoplanet compositions.  We would also like to thank Glenn Wolfgang for his thoughts and perspective on this project as a formally trained statistician.  The authors' financial support during this investigation was primarily provided by the University of California Eugene Cota-Robles Graduate Fellowship (AW) and by the NASA Ames NAI Astrobiology Module (GL).  This material is also based upon work supported by the National Science Foundation Graduate Research Fellowship under Grant No. 0809125.

Most of the data presented in this paper were obtained from the Multimission Archive at the Space Telescope Science Institute (MAST). STScI is operated by the Association of Universities for Research in Astronomy, Inc., under NASA contract NAS5-26555. Support for MAST for non-HST data is provided by the NASA Office of Space Science via grant NNX09AF08G and by other grants and contracts.  This research has also made use of the Exoplanet Orbit Database and the Exoplanet Data Explorer at exoplanets.org.



{\it Facilities:} \facility{Kepler}.




\clearpage

\end{document}